\documentclass[article,10pt,twocolumn]{Nicks}
\usepackage{mathtools}
\usepackage{graphicx}
\usepackage{xcolor}
\usepackage{multicol}
\usepackage{wrapfig}
\usepackage{lipsum}  
\usepackage{dcolumn}
\usepackage{braket}
\usepackage{appendix}
\usepackage{hyperref} 
\usepackage{amsmath,amsthm,amssymb,amsfonts}
\usepackage{bm}
\usepackage[utf8]{inputenc}
\usepackage[acronym]{glossaries}
\usepackage{subcaption}
\usepackage{pdfpages}
\usepackage{afterpage}

\usepackage{xcolor}
\usepackage{graphicx}
\usepackage{float}
\usepackage{mhchem}
\usepackage{tabularx}

\usepackage[utf8]{inputenc}
\usepackage[T1]{fontenc}
\usepackage{enumitem}
\usepackage{subfiles} 
\title{Topological and conventional nano-photonic waveguides for chiral integrated quantum optics}

\author[1,*,+]{N.J Martin}
\author[2,+]{M. Jalali Mehrabad}
\author[1]{X. Chen}
\author[1]{R. Dost}
\author[3]{E. Nussbaum}
\author[1]{D. Hallett}
\author[1]{L. Hallacy}
\author[1]{A. Foster}
\author[4]{E. Clarke}
\author[4]{P.K. Patil}
\author[3]{S. Hughes}
\author[2]{M. Hafezi}
\author[1]{A.M Fox}
\author[1]{M.S.~Skolnick}
\author[1]{L.R. Wilson}

\affil[1]{Department of Physics and Astronomy, University of Sheffield, Sheffield S3 7RH, UK}
\affil[2]{Joint Quantum Institute, University of Maryland, College Park, MD 20742, USA}
\affil[3]{Centre for Nanophotonics, Department of Physics, Engineering Physics and Astronomy, Queen’s University, Kingston, Ontario, Canada K7L 3N6}
\affil[4]{EPSRC National Epitaxy Facility, University of Sheffield, Sheffield S1 4DE, UK}

\affil[*]{nmartin3@sheffield.ac.uk}
\affil[+]{these authors contributed equally to this work}

\begin{document}

\date{\today}

\begin{abstract}

Chirality in integrated quantum photonics has emerged as a promising route towards achieving scalable quantum technologies with non-linearities at the single photon level. Topological photonic waveguides have been proposed as a novel approach to harnessing such chiral light-matter interactions on-chip. However, uncertainties remain regarding the strength of the chiral coupling of embedded quantum emitters to topological waveguides in comparison to conventional line defect waveguides. In this work, we present an investigation of chiral coupling in a range of waveguides using a combination of experimental, theoretical, and numerical analyses. We quantitatively characterize the position-dependence of the light-matter coupling on several topological photonic waveguides, and benchmark their chiral coupling performance against conventional line defect waveguides. We conclude that topological waveguides under perform, in comparison to conventional line defect waveguides, casting their directional optics credentials into doubt. To demonstrate this is not a question of the maturity of the field, we show that state of the art inverse design methods, while capable of improving the directional emission of these topological waveguides, still places them significantly behind the operation of a conventional (glide-plane) photonic crystal waveguide. Our results and conclusions pave the way towards improving the implementation of quantitatively-predicted quantum nonlinear effects on-chip.

\end{abstract}

\maketitle 

\section{Introduction}

Integrated nano-photonic platforms, in which embedded quantum emitters are interfaced with optical waveguides and cavities on-chip, are a promising route towards scalable quantum technologies. An attractive property of nanophotonic waveguides is their support for chiral light-matter interactions, whereby an emitter with a circularly polarised transition dipole moment couples unidirectionally at the single photon level to a single photonic waveguide mode \cite{Lodahl2017,Sollner2015,Coles2016,polar}. Such interactions have previously been demonstrated on-chip using semiconductor quantum dots (\textsc{QD}s) coupled to photonic crystal (PhC) line defects such as W1 \cite{Coles2016} and glide plane \cite{Sollner2015,lodahl_glide_plane,Hamidreza_glide} waveguides. 

Recently, PhC topological waveguides have received significant interest for integrated nanophotonics due to their attractive properties, which include robust transmission around tight bends \cite{Barik666,Ma_2019,Yamaguchi_2019,He2019,Shalaev2019,Parappuratheaaw4137,Yoshimi:21,MJmehrabad2021,mehrabad2023topological}, opening up their applications to more complex guided structures, such as ring resonators, beam splitters and filters that require sharp bends \cite{Barik_2020,JalaliMehrabad_APL,JalaliMehrabad_Optica,Gu_2021}. 

\color{black}More pertinently for chiral quantum optics, topological waveguides were envisioned to have two key properties, 1. robustness to in-plane back-scattering and 2. intrinsic unidirectional emission for embedded emitters, with the unidirectional emission expected to arise from the intrinsically helical edge modes which arise at the interface between two topologically-distinct PhCs \cite{Barik666,Mehrabad2023,mehrabad2023topological}.

Despite intense research on the application of these systems, experimental measurement of the degree of robustness to back-scattering of these waveguides was only very recently examined using external light sources \cite{backscattering}, the results casting considerable doubt to the resistance to back scattering. However, an experimental investigation into the envisioned property of intrinsic unidirectional emission of topological waveguides, key to their chiral quantum optics applications, remains missing.

\color{black} 
\begin{figure*}[p]

\includegraphics[width=\textwidth]{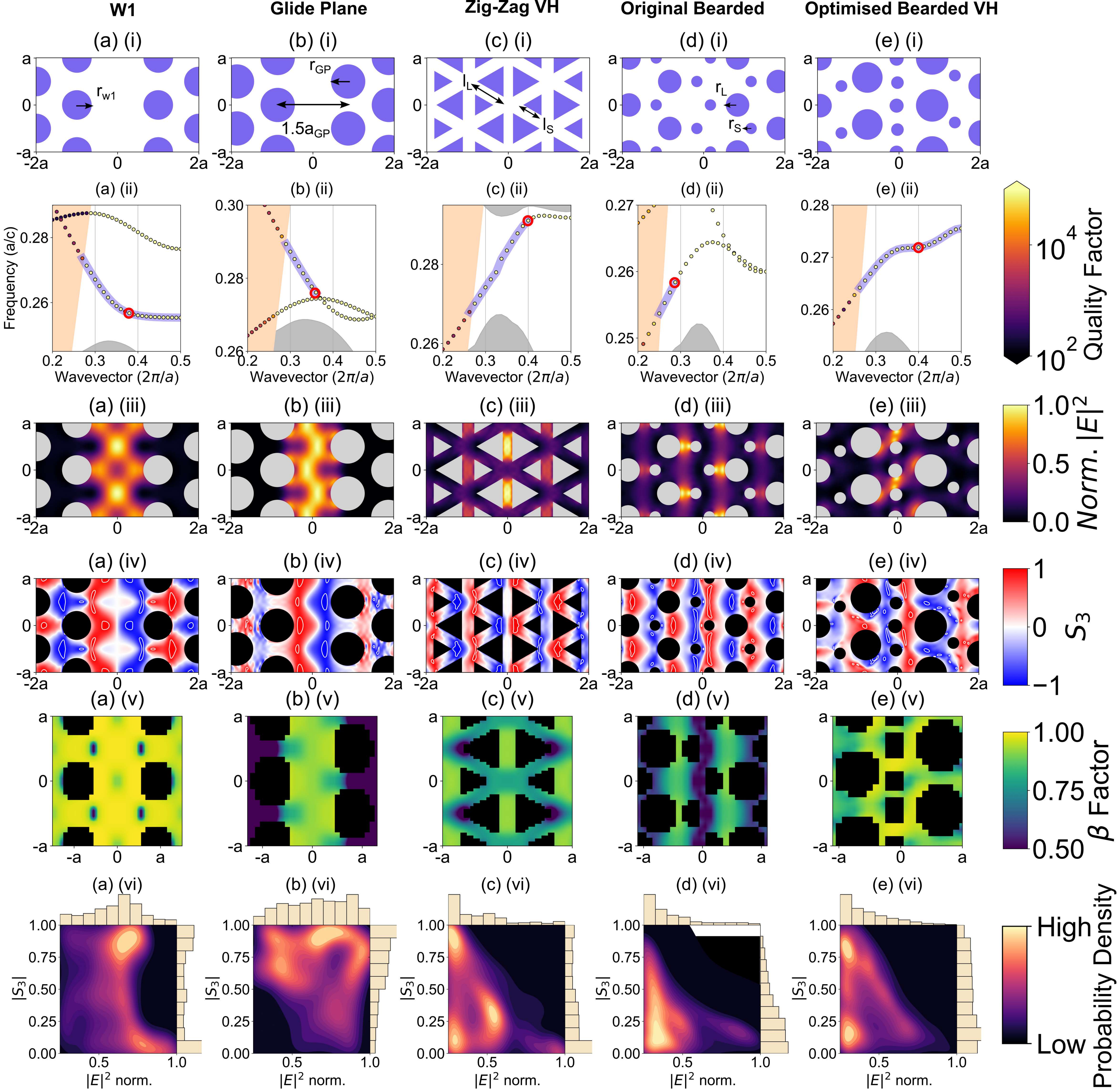}

\caption{(a-e) (i), schematics of the (a) W1 waveguide, (b) Glide plane waveguide, (c) Zig-Zag interface VH topological waveguide, (d) Un-optimised Bearded interface VH topological waveguide and (e) a Bearded interface VH topological waveguide optimised using an inverse design technique for a more favourable band structure and improved electric field and $S_{3}$ overlap. (a-e) (ii) simulated band structures of the waveguides, with the single mode region of interest highlighted in blue, and the specific frequency of single mode operation chosen for the electric field plots in the rest of the figure highlighted on the band structures in red. (a-e) (iii) Simulated electric field profiles and (a-e) (iv) $S_3$ maps for the waveguides. Within (a-e) (iv), encircled white regions indicate regions where chiral contrast is expected to be 95\% and above. Band structures and electric field plots were simulated by guided mode expansion. (a-e) (v) Shows the spatial dependence within the waveguides of the $\beta$-factor at the same frequency as the electric field plots, simulated using FDTD. Black regions indicate the position of the waveguide apertures. (a-e) (vi) Probability density plots showing the likelihood of a set of randomly positioned dots possessing a given $S_{3}$ value for different electric field strengths within the waveguide interface. Areas of high probability density indicate combinations of chiral contrasts and electric field strengths that are more likely. The apertures of the waveguides are excluded from the calculations.  \label{Figfield} }

\end{figure*}

As in conventional, non-topological waveguides, chiral coupling of an embedded emitter to a topological photonic waveguide is position dependent. In the most extreme case, the direction of emission from a circularly polarised transition can be completely reversed by moving it within a unit cell of the PhC lattice. While the properties of conventional waveguides and topological waveguides have been compared in simulations \cite{lodahl_hughes}, here we use both simulations and experiment to compare these  approaches to realising chiral light-matter interactions on-chip.

Within the main text we present data on conventional W1, glide plane, zig-zag interface valley-Hall (VH) and bearded interface VH waveguides. In addition we include a novel version of the bearded interface VH waveguide, optimised using inverse design techniques. (See supplementary information for a discussion of slab waveguides and an optimised glide plane waveguide\cite{Hamidreza_glide} waveguides) For each waveguide we present the dispersion relation of the guided mode, the spatial electric field intensity and degree of circular polarisation as well as the spatial and wavelength dependence of the $\beta$-factor (see section 3.1), for emitters within the waveguides. We use this simulated data to make predictions for the expected distribution of the chiral contrast of embedded QDs and then determine experimentally the chiral contrast for a large number of QDs in each type of structure, showing good agreement with the simulations. While the topological waveguides, optimised through inverse design, demonstrate promising improvements in chiral properties in simulation, our experimental results also serve to highlight the limitations of current approaches. In particular we stress the need to develop waveguides which support high $\beta$ factors whilst simultaneously showing near-unity chiral coupling.

\section{Photonic crystal waveguide designs for chiral quantum optics}

The PhC waveguides considered in this work are described schematically in Fig 1 (a-e). We consider two conventional (i.e. non-topological) and three non-trivial topological designs. For each type of waveguide we show:
\begin{enumerate}[label=\roman*]
    \item The waveguide design.
    \item The band structure.
    \item The electric field profile.
    \item The chirality ($S_{3}$).
    \item The $\beta$-factor.
    \item $S_{3}$ and E-Field probability density plots.    
\end{enumerate}

For each type of waveguide we focus on the parts of the band structure shaded in blue,
and the field profiles in panel (iii-v) are calculated for the points shown by the red circle in panel (ii). The first  conventional design (Shown in Fig \ref{Figfield}(a)) is that of a W1 waveguide, comprising a triangular lattice of circular holes etched into a thin dielectric membrane, with one row of holes omitted in the $\Gamma$-K direction to form a line defect. Shown in Fig \ref{Figfield} (a) (ii), several guided Bloch modes can be observed within the PhC bandgap of the W1 waveguide; we focus on the lowest frequency mode, with the field profile shown in Fig \ref{Figfield} (a) (iii), which has an electric field antinode at the centre of the waveguide.

The second conventional design studied here is the glide plane waveguide, which is formed by displacing the holes on one side of a W1 waveguide by half the lattice period along the waveguide.  The TE like mode dispersion diagram for the glide plane structure is shown in Fig \ref{Figfield} (b) (i). We focus on the single mode region of the higher frequency mode, as the multimode region where modes the two modes overlap spectrally prevents chiral coupling being realised \cite{Mahmoodian:17}. The field profile for this mode is shown in Fig \ref{Figfield} (b) (iii).

The conventional waveguides introduced above are compared in this work with three topologically non-trivial structures. We have chosen valley-Hall waveguides as their interfaces support guided optical modes lying below the light line, unlike the alternative spin-Hall approach \cite{Barik666,JalaliMehrabad_APL,PhysRevResearch.2.043109}. VH waveguides can be formed by interfacing two VH photonic crystals where the unit cell is a rhombus containing a pair of apertures of differing sizes, in two distinct ways. The first, the zig-zag interface, is an interface of these two crystals with mirror symmetry. For this work, a design for the zig-zag interface comprised of triangular apertures, with the larger triangles at the interface. This design is chosen for its favourable band structure, with the dispersion diagram in Fig \ref{Figfield} (c) (ii) showing that the waveguide supports a single guided TE mode, with the mode profile given in Fig \ref{Figfield} (c) (iii). The bearded interface in contrast, is formed by interfacing two valley hall photonic crystals such that it forms a glide-plane symmetric interface. For this work, the original bearded interface waveguide is formed of circular apertures, with the smaller of the apertures at the interface. We have also included an optimised version of the bearded interface within our investigation. Here, an inverse design algorithm was used to optimize the geometrical parameters of a bearded interface VH nano-photonic waveguide for high $\beta$-factors and strong chiral light-matter interactions (see supplementary section S1 for more details on the inverse design process) 


\section{Simulation of the waveguide $\beta$-factor, E-field properties and chirality}

\begin{figure}[t]
 \centering
 \includegraphics[width=0.47\textwidth]{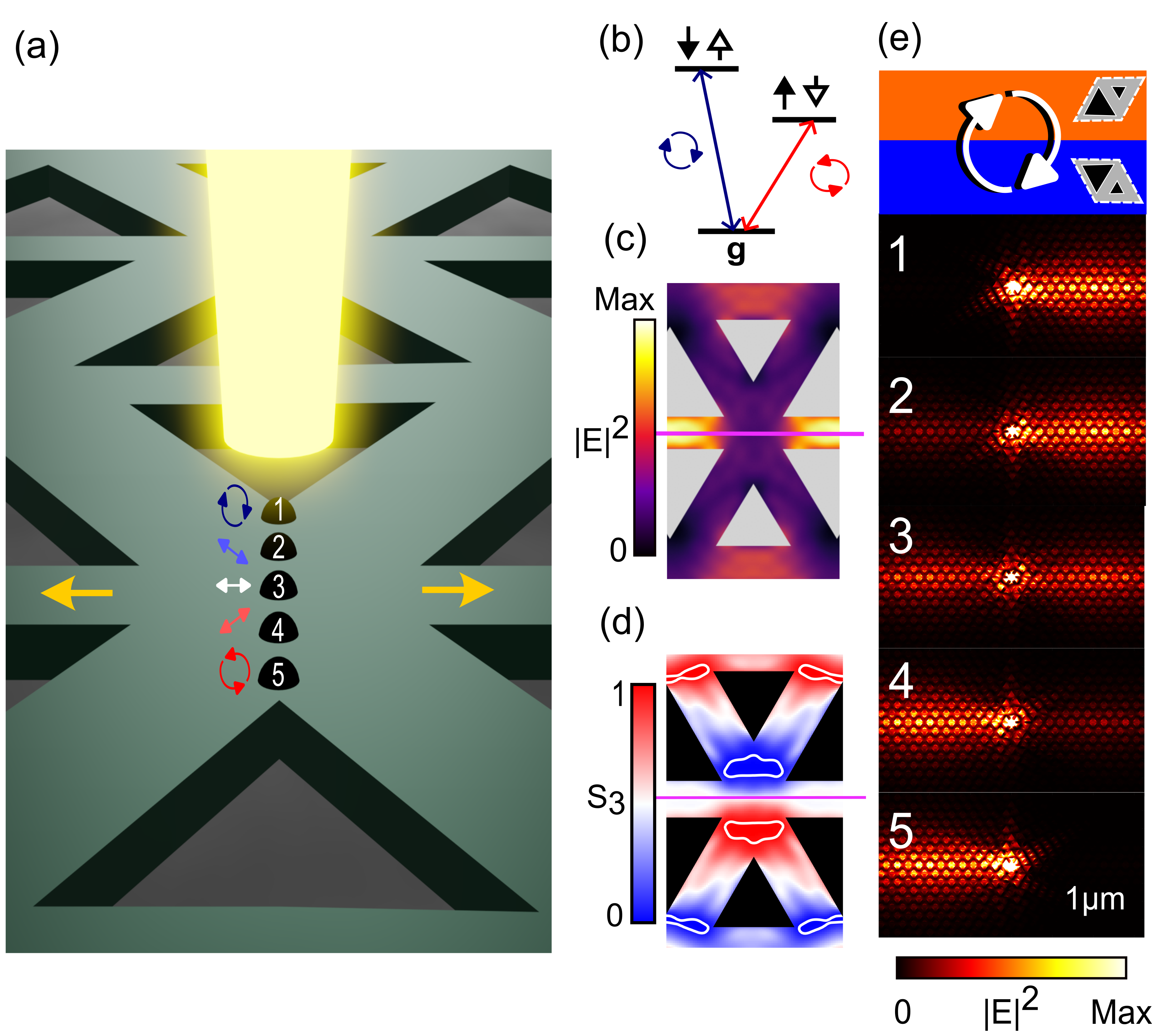}
    \caption{  (a) Schematic diagram showing the positional dependence of the direction of emission of a circularly polarised dipole source within a VH waveguide; labels show the polarisation of the local field at the different positions, with the waveguide interface highlighted in magenta. (b) Level diagrams and optical transitions for an neutral exciton QD state showing left and right circularly polarised transitions.  (c) $|E|^{2}$ field profile within the VH waveguide interface. (d) $S_{3}$, degree of circular polarisation at different points within the VH waveguide interface. Encircled white regions indicate regions where $|S_{3}| \geq 0.95$ (e) FDTD simulations showing the positional dependence of emission from a circularly polarised dipole at points 1-5 within the ziz-zag VH waveguide interface. The top of the figure shows the position of the waveguide interface between two topologically distinct photonic crystals (shown in orange and blue) as well as the central position of the right circularly polarised dipole source used in the simulations. }
    \label{Chiral couple}
\end{figure}

This section presents simulation results, that quantify the spatial dependence of the $\beta$-factor, E-field and chiral contrast. This includes a description of the model used to predict the chiral statistics i.e. the expected statistical distribution of chiral contrast for a large number of randomly positioned QDs.

\subsection{$\beta$-Factor}

We define our $\beta$-factor as the ratio of the spontaneous emission rate of an emitter into an intended guided mode compared to the total decay rate. To calculate the $\beta$-factor we used finite-difference-time-domain (FDTD) simulations to calculate the fraction of radiative power coupled into propagating modes over the total power injected from the emitter. To determine the spatial distribution of the $\beta$-factor, we vary the dipole position within a region from approximately -2a
to +2a from the waveguide centre, excluding regions within the apertures of the waveguides (see supplementary section S3-5 for more information on the calculation of the $\beta$ factor as well as its spectral dependence). 

\color{black}
The relationship between the intensity of the electric field at a given point to the $\beta$-factor at that point can be seen by comparing the electric field profiles of Fig 1 (a-e) (iii), and the $\beta$-factors of Fig 1 (a-e) (v), with areas of high electric field intensity resulting in higher $\beta$-factors. However, the enhancement of the $\beta$-factor arising from the slow light regions of the waveguides can result in regions with relatively low electric field intensities having high $\beta$-factors, with this effect most notable within the W1. This wavelength dependence of the $\beta$-factor is shown within the SI and offers waveguides with poor E-field and $S_{3}$ overlap a route to high $\beta$-factor at chiral points.

\subsection{Stokes $S_{3}$ parameter and chiral contrast}

The chiral properties of each waveguide were evaluated using guided mode expansion simulations \cite{Minkov2020} by calculating the Stokes $S_3$ parameter defined as:
\begin{equation}
    S_3=\frac{-2\textrm{Im}(E_xE_y^{*})}{\lvert E_x\rvert^2+\lvert E_y\rvert^2}.
    \label{s3}
\end{equation}
The simulated Stokes $S_{3}$ parameter is shown in Fig \ref{Figfield} (a-e) (iv) for the five waveguides. The Stokes parameter at the position of a QD determines the chiral contrast of the emission. Fig \ref{Chiral couple} shows the relationship between the $S_{3}$ parameter and the direction of emission of a circularly polarised emitter, such as the Zeeman split lines of a neutral excition within a QD (Fig \ref{Chiral couple} b). The chiral contrast is defined as: 
\begin{equation}
     C = \frac{T^{R} - T^{L} }{T^{R} + T^{L}}, 
     \label{chiral}
\end{equation}
where $T^{R}$ and $T^{L}$ are the transmission right and left through a waveguide respectively for an emitter embedded within a linear waveguide. Within the FDTD simualtions of Fig \ref{Chiral couple}, the chiral coupling of a circularly polarised emitter (within the example of the VH zig-zag interface) is shown to be position dependent, due to the different $S_{3}$ values at different points within the waveguide. While all five waveguides show regions of high chiral contrast, with contrasts above 95\% indicated by the in-circled white regions of Fig (a-e) (iv), S3 does not give the complete picture of a waveguide’s suitability to chiral quantum optics, as discussed in the next section.

\subsection{Coincidence of high chirality and E-field concentration}

The likelihood of a set of randomly positioned QDs possessing a given $S_{3}$ value for different
electric field strengths within the waveguide interface is shown by the probability density
plots of Fig (a-e) (iv). Areas of high probability density indicate combinations of chiral contrasts and electric field strengths that are more likely; the air regions of the waveguides where there are no QDs are excluded from the calculations. Within these plots we see that the glide plane waveguide has the best overlap of $S_{3}$ and $|E|^{2}$. The W1 waveguide shows a high concentration of QDs at high $S_{3}$ for $|E|^{2}=0.7$, which outperforms the distributions seen for the topological waveguides, characterised by low density at the intersection of high $S_{3}$ and $|E|^{2}$ found in the dark regions at the top right of the plots. The optimised bearded interface does, however, perform better than the un-optimised version in this regard, with the optimised waveguide showing a greater concentration of high chirality QDs at $|E|^{2}=0.3$.

\begin{figure*}[t!]
\includegraphics[width=0.99\textwidth]{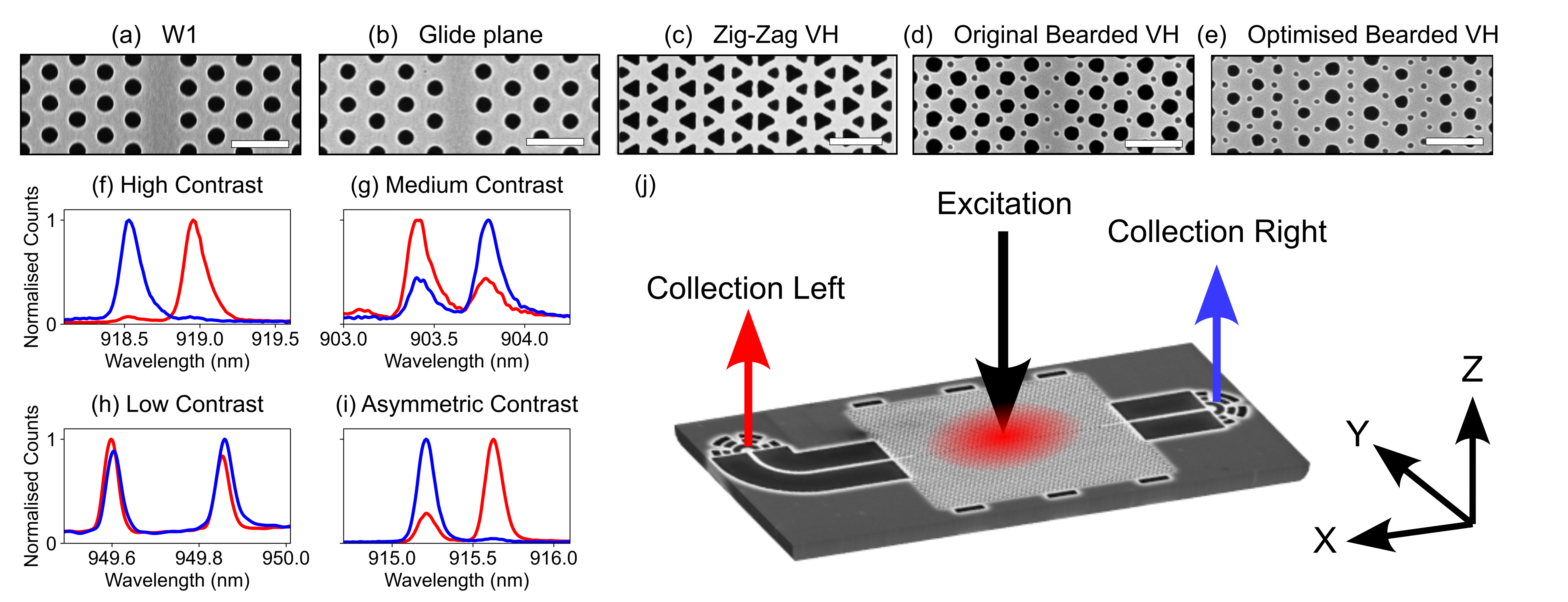}
\caption{SEM images of the fabricated waveguide for (a) W1, (b) Glide plane, (c) zig-zag VH, (d) original bearded VH and (e) optimised bearded VH. Scale bars are 0.4$\mu m$. Examples of dot lines split by B$\neq$0 magnetic fields, with (f) high chiral contrast, (g) medium contrast, (h) low contrast  and (i) asymmetric contrast. (j) Schematic of the on chip device layout, indicating the location of the left out-coupler collection (red) and the right out-coupler collection (blue) \label{Figsem} }
\end{figure*}

\subsection{Simulated Chiral Statistics}

The chiral statistics of the waveguides can be modelled by considering the $S_3$ field maps for
the waveguides over the selected waveguide regions highlighted in blue in Fig 1 (a-e) (ii).
By calculating the electric field data using guided mode expansion for the middle of the slab (where the QDs are located) for multiple points within a 10a super cell of the waveguide, the distribution of the chirality can be calculated. With FDTD simulations providing data for the $\beta$-factor (see Fig \ref{Figfield} (a-e) (v)) and Purcell factor, a threshold was applied to the points that were included within the chiral statistics. This threshold was set so that only points within the waveguide that had a combined value for Purcell factor multiplied by $\beta$-factor above 0.5 were included ( $F_{P}\cdot\beta \geq 0.5$). Additionally, to account for the `dead-zones' for QD emission that exist around the apertures of the waveguides. The extend of this dead-zone can depend on the wafer properties and the presence of fabrication processes such as surface passivisation \cite{MANNA2020147360}, for this reason we have calculated the expected statistics for the waveguides excluding a region of both 15 and 30 nm \cite{Pregnolato} around the apertures. The absence of QD emission in proximity to the etched surfaces is likely to arise from detrimental effects of surface defects and interface roughness. The simulations of the chiral statistics for the five types of waveguides are shown in Fig 4(a)-(e). These results will be discussed in the next section when comparing to the experimental data.
\color{black}


\begin{figure*}[t]
     \centering

     \begin{subfigure}[b]{0.99\textwidth}
         \centering
         \textbf{\large \sffamily}\par\medskip         
         \includegraphics[width=\textwidth]{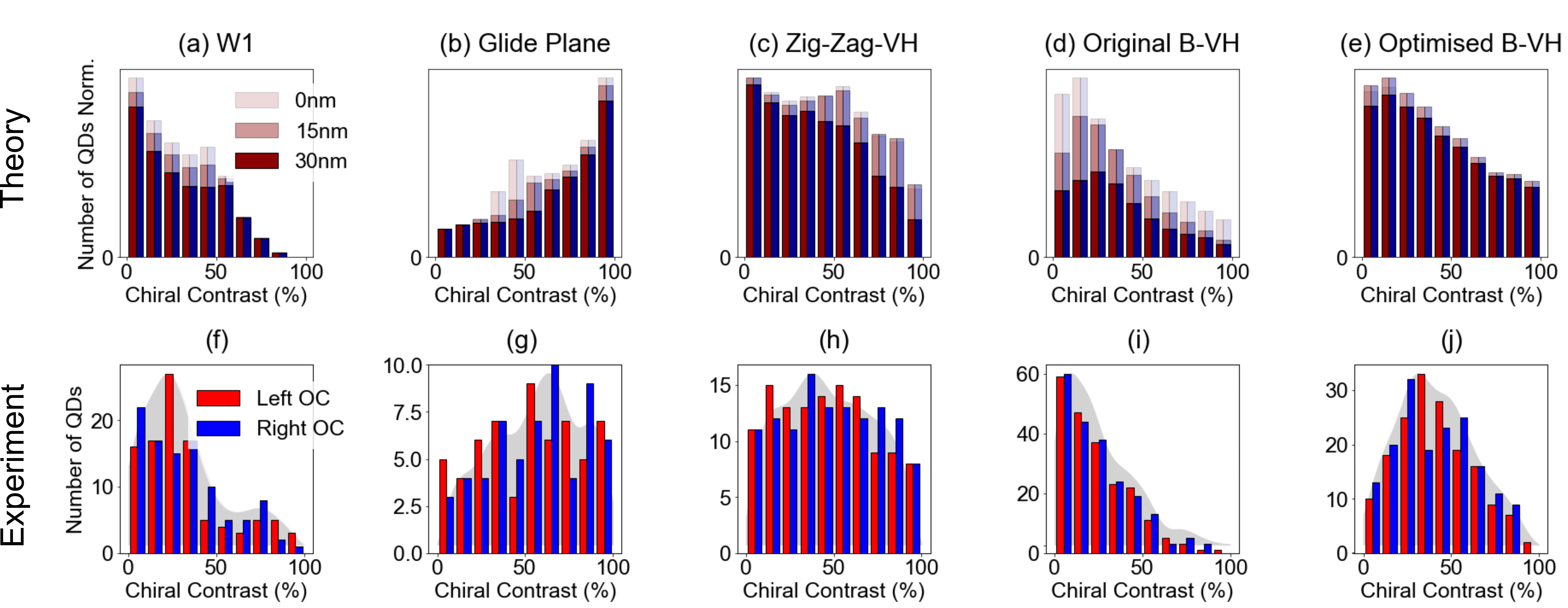}

     \end{subfigure}

        \caption{(a-e) Modeled predictions for the chiral statistics of the waveguides for a 0nm and 15nm and 30nm dead-zone with the plots normalised to the 0nm case to show the reduction in the expected number of dots that arise from these dead-zones. (f-j) Chiral contrast measured experimentally for randomly positioned QDs within the different waveguides. Left out-coupler data is presented in red and Right out-coupler data in blue. All experimental data were recorded when an external magnetic field of $B_Z$=3T was applied. The grey background plot shows a smooth distribution of the chirality data where no distinction between the results of the left and right out-couplers are made.}
        \label{fig_results}
\end{figure*}

\section{Method}
\subsection{Experimental comparison of the photonic crystal waveguides}


The experimental results were obtained for devices fabricated in a ~175nm-thin GaAs \textit{p-i-n} membrane containing a single layer of InAs QDs. Representative scanning electron microscope (SEM) images of the waveguide interfaces are shown in Fig \ref{Figsem} (a-e). Each waveguide was terminated on both ends with a grating coupler for light extraction into external optics, as shown in \ref{Figsem} (j). The sample was cooled to 4.2K in a superconducting magnet cryostat. To determine the operation bandwidth of each waveguide, broadband photoluminescence (PL) was generated by exciting the ensemble of QDs located within one grating coupler using non-resonant excitation ($\lambda_{\textrm{laser}}=810$nm). PL emission was then detected from the other outcoupler, with representative transmission spectra shown in SI Fig S2. A combination of the identification of a sharp decrease in transmission resulting from the band edge of a guided mode and an analysis of the Fabry-P\'{e}rot fringes of the waveguide were used to identify the single mode regions of interest. (More information on the identification of the single mode regions of the waveguides can be found in the supplementary section S2).

Next, we measured the chiral contrast independently for a large number of individual QDs in each waveguide. To do so, we used low power micro-photoluminescence ($\mu\mathrm{PL}$) measurements, exciting non-resonantly from above the waveguide and collecting emission independently from the two out-couplers. We focused on QDs spectrally located in the single mode regions of the waveguides, highlighted in blue in Fig \ref{Figfield} (a-e). In the presence of a Faraday-geometry magnetic field, the circularly polarised dipole transitions of the QD split energetically, allowing for their spectral resolution. To allow for the possibility of dissimilar out-coupler collection efficiency, the chiral contrast was evaluated independently for each out-coupler using the relationship:

\begin{equation}
C_i = \frac{I_i^{\sigma^{+}}-I_i^{\sigma^{-}}}{I_i^{\sigma^{+}}+I_i^{\sigma^{-}}},
\end{equation}

where $I_i^{\sigma^j}$ ($j=+,-$ ) represents the intensity of $\sigma^j$ polarised light emitted by the QD and collected from OC $i$ (= left, right).
\subsection{Experimental chiral statistics}

We quantify the degree of chiral coupling in each nano-photonic waveguide experimentally, using randomly distributed self-assembled InAs QDs, distributed by growth in the $xy-$plane at $z = 0$ of the GaAs membrane (see Fig 3 (j) for axes). The guided PL of single QDs is detected from right and left out-couplers (denoted as Left OC and Right OC in Fig \ref{fig_results} (f-j)) at opposite ends of each waveguide. Self-assembled growth leads to random positions of QDs within the xy-plane, which leads to coupling of QDs placed at areas of the waveguide with a range of chiral contrast. The results are shown in histogram graphs for the experimentally measured chiral contrasts for each case. The binned histograms from the $S_3$ parameters calculated via FDTD simulations, as discussed in section 3.4, are also shown for comparison.


For the W1 waveguide we see that, in general, there is predicted to be a low proportion of QDS with a chirality > 80\%, but a high proportion of dots with low chiral contrast. This can be explained by both the concentration of the electric field being at points of low circular polarisation, as seen in Fig \ref{Figfield} (f,k), and the annihilation of chiral points at the band-edge in a W1 waveguide \cite{chiral_points}. The experimental results broadly agree with this prediction.

The glide-plane PhC waveguide exhibits the best chiral coupling for QDs in both simulation and experiment. Nevertheless there is a lower proportion of high contrast QDs in experiment than expected. This may be explained by the glide plane's high chiral contrast regions being located near the waveguides' etched areas.

The topological VH Zig-Zag interface achieves a high proportion of QDs with high contrast, but not as high as the glide plane, and as can be seen in Fig \ref{Figfield} (e) (vi), these high contrast dots are unlikely to be at points of high electric field concentration. The original bearded interface, however, while predicted to have a better distribution than the W1 waveguide, performs the worst out of all of the waveguides in experiment. This is likely related to the interface holes in the design being a source of fabrication error due to their very small size, and also to the surface proximity issues that they introduce beyond those included in our model. The optimised bearded VH waveguide shows a significant improvement in the experimental results in comparison to its un-optimised counterpart, but not the improvement implied by the simulation data. This is likely due a combination of two factors. The first is the high group indexes of the ideal design, not being replicated in experiment, resulting in a lower probability of high contrast QDs being measured. 
The second is the interface holes again leading to a reduction in the contrast for reasons described above. 

\color{black}
A feature that is present within the statistics for all of the waveguides, and has been seen in other experiments and QD wafers \cite{Coles2016}, is the asymmetry in the chirality of the QDs measured from the left and right out-couplers of the waveguide. Fig 3 (f-i) shows examples of QD spectra that fit closely with the predictions of the FDTD simulations shown in Fig 1 (a). These spectra show the expected behaviour of the spectra of the right out-coupler (red) and the left out-coupler (blue) are symmetric, mirrored versions of each other, displaying the same chiral contrast with opposite intensities. However, Fig 3 (f) shows an example of asymmetric behaviour where the contrast as viewed from the left out-coupler is greater than is seen in the right out-coupler. (A discussion on the extent and nature of this asymmetry can be found in the SI section S6).

\color{black}

\section{Conclusion}
 \color{black}
We conducted a comparative analysis of conventional and topological waveguides for chiral coupling of embedded quantum emitters. Through a combination of experimental characterization of chiral coupling and a simulation-based analysis of electric field properties and $\beta$-factors, we have gained a comprehensive view of the waveguides' individual suitability for chiral quantum optics applications. Among the waveguides examined, the glide plane demonstrated the highest proportion of high chiral contrast QDs, making it well-suited for achieving such QDs within a linear waveguide system with randomly positioned QDs. In contrast, while the zig-zag interface VH waveguide exhibited good experimental performance by producing a large portion of high chiral contrast QDs, our modelling showed that regions with high chiral contrast within the zig-zag waveguide are unlikely to have high $\beta$-factors.

When considering more complex structures like ring resonators that require robust transmission around tight bends, topological waveguides emerge as a promising platform for chiral quantum optics. Despite the discussed $\beta$-factor limitations of the zig-zag interface, it is still possible to achieve high $\beta$-factor within topological waveguides by using slow light. By combining the high $\beta$-factors resulting from the slow light region of the inverse designed bearded VH waveguide with appropriate QD registration \cite{Coles2016, Sapienza2015, Schnauber, Pregnolato} or site-controlled growth \cite{site_control,Site-Controlled_jons} techniques, it may be possible to achieve deterministic positioning of QDs at points within the waveguide that exhibit both high chirality and high $\beta$-factors. This approach presents a pathway towards achieving high beta-factor, high chiral contrast QD emission within more complex photonic crystal device geometries.

Exciting future prospects of this research include the realization of separation-independent QD-QD interactions \cite{Lodahl2017,PhysRevB.92.155304}, super and subradiant many-body states \cite{science_lodahl,edo_super,Grim2019}, and the formation of large-scale chiral spin networks \cite{Pichler} using a conventional or topologically-protected photonic platform. Spin \cite{suarez2023spin} and vorticity \cite{session2023optical} selective light-matter coupling in the quantum Hall regime is also an intriguing new platform, in which the topological interplay between light and matter can be explored.

\color{black}
\section*{Acknowledgements }

This work was supported by EPSRC Grant No. EP/N031776/1, EP/V026496/1, the Quantum Communications Hub EP/T001011/1 and the natural sciences and engineering council Canada (NSERC). The authors would like to acknowledge helpful discussions with Edo Waks, Kartik Srinivasan, Nir Rotenberg, Peter Millington-Hotze and Hamedreza Siampour. 

\section*{Author contributions statement}
N.J.M., M.J.M., and E.N. designed the photonic structures, which R.D. fabricated. E.C. and P.K.P. grew the sample. N.J.M., M.J.M., X.C., L.H., A.F. and  D.H. carried out the measurements and simulations. L.R.W., A.M.F, S.H., M.H., and M.S.S. provided supervision and expertise. N.J.M., M.J.M., X.C. L.H. and E.N. wrote the manuscript, with input from all authors.

\section*{Data Availability}
Data supporting this study are openly available from the authors upon reasonable request.


\bibliography{Bibli}
\newpage
\onecolumn
\begin{center}
\textbf{\Huge \textsf{Supplementary Materials}}
\end{center}
\setcounter{equation}{0}
\setcounter{figure}{0}
\setcounter{table}{0}
\setcounter{page}{1}
\setcounter{section}{0}
\makeatletter
\renewcommand{\thepage}{S\arabic{page}}
\renewcommand{\thesection}{S\arabic{section}}
\renewcommand{\thefigure}{S\arabic{figure}}
\renewcommand{\theequation}{S\arabic{equation}}

\section{Waveguide Schematics and optimisation}

\begin{figure*}[h]
\centering
\includegraphics[width=0.99\textwidth]{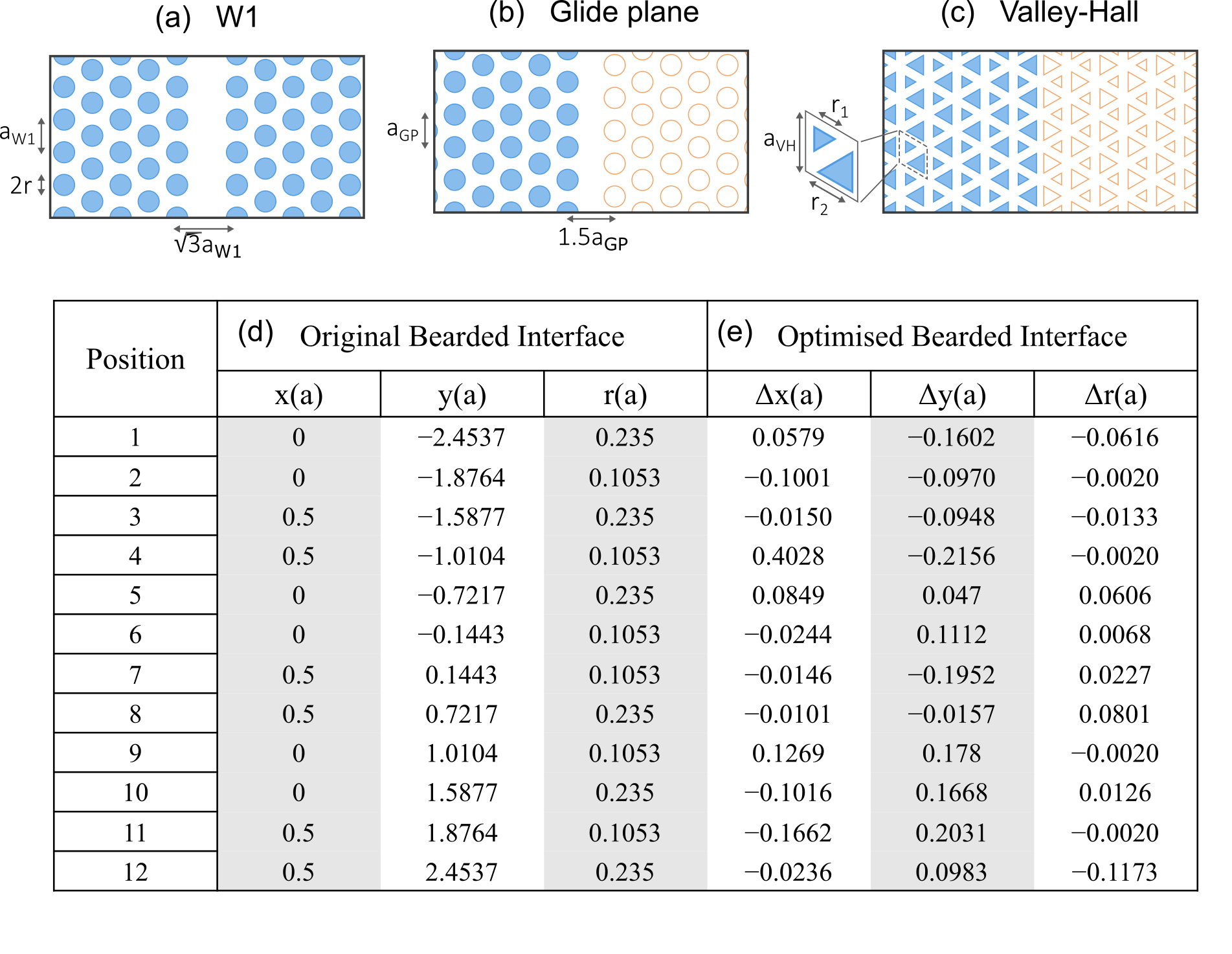}
\caption{Schematics and parameters for the waveguides used in the main text. (a) Schematic for the W1 waveguide where $t=0.763a_{\mathrm{W1}}$ and $r=0.3a_{\mathrm{W1}}$. (b) Schematic for the glide-plane waveguide, where $t=0.69a_{\mathrm{GP}}$ and $r=0.35a_{\mathrm{GP}}$. (c) Schematic for the zig-zag VH waveguide, where $t=0.64a_{\mathrm{VH}}$, $r_1=0.4a_{\mathrm{VH}}$ and $r_2=0.6a_{\mathrm{VH}}$. (d) Location and radius of the 12 holes, closest to the interface in the initial bearded interface VH waveguide design. (e) Location and radius of the 12 holes, closest to the interface in the optimised bearded interface VH waveguide design in terms of changes relative to the original waveguide. A dielectric constant of $\epsilon=11.6$ was used throughout.\label{Fig2} }
\end{figure*}

To optimise the bearded interface waveguide, an inverse design technique is used, comprised of Figures of merit (FOM) used in combination to yield more favourable properties for the topologically non-trivial mode including higher group indexes within a single mode region, and better overlap of the electric field and chiral points. A two step optimization is run to obtain the improved design. In the first step $\mathcal{F}_1$ is maximized and in the second step $\mathcal{F}_2$ is minimized. $\mathcal{F}_1$, is:
\begin{equation}
\mathcal{F}_1={\rm max}(G_SM\cdot C \cdot F_p^R) F_{\rm SM},
        \label{eq: FOM}
\end{equation}
where $G_SM$ is a smooth step function (equal to 0 inside a hole edge and equal to 1 more than 40~nm from the closest hole edge), $C$ is the chiral contrast, $F_p^R = \Gamma^f/\Gamma^{\hom}$ is the Purcell factor, and $F_{\rm SM}$ is a function that encourages a certain magnitude and position of single mode bandwidth. The second phase of the optimization is focused on improving the dispersion, with a figure of merit, $\mathcal{F}_2$, of:
\begin{equation}
    \mathcal{F}_2=\omega_{bb} - \omega_{{\rm SM}, \max} + \frac{A}{\max_{\rm SM}(n_g)},
\end{equation}
where $\omega_{bb}$ is the frequency where the topologically non-trivial band first bends back, else, in the absence of a bend, it is the frequency of the topologically non-trivial band at $k_x a = \pi$. The parameter $\omega_{{\rm SM}, \max}$ is the maximum frequency of the topologically non-trivial mode's lowest frequency single mode bandwidth, $A$ is a weight to balance the third term, and $\max_{\rm SM}(n_g)$ is the maximum $n_g$ in the topologically non-trivial mode's lowest frequency single mode bandwidth. This FOM works to eliminate any bending back of the topologically non-trivial mode and to increase the maximum single mode group index. The improved guided mode group index can be seen in the flatter band of Fig 1 (e) (ii) in comparison to (d) ii. More information about this optimisation and design of the optimised bearded interface waveguide can be found in Ref \cite{PhysRevA.106.033514}.

\newpage

\section{Identification of Waveguide Regions}

\begin{figure*}[h]
\centering
\includegraphics{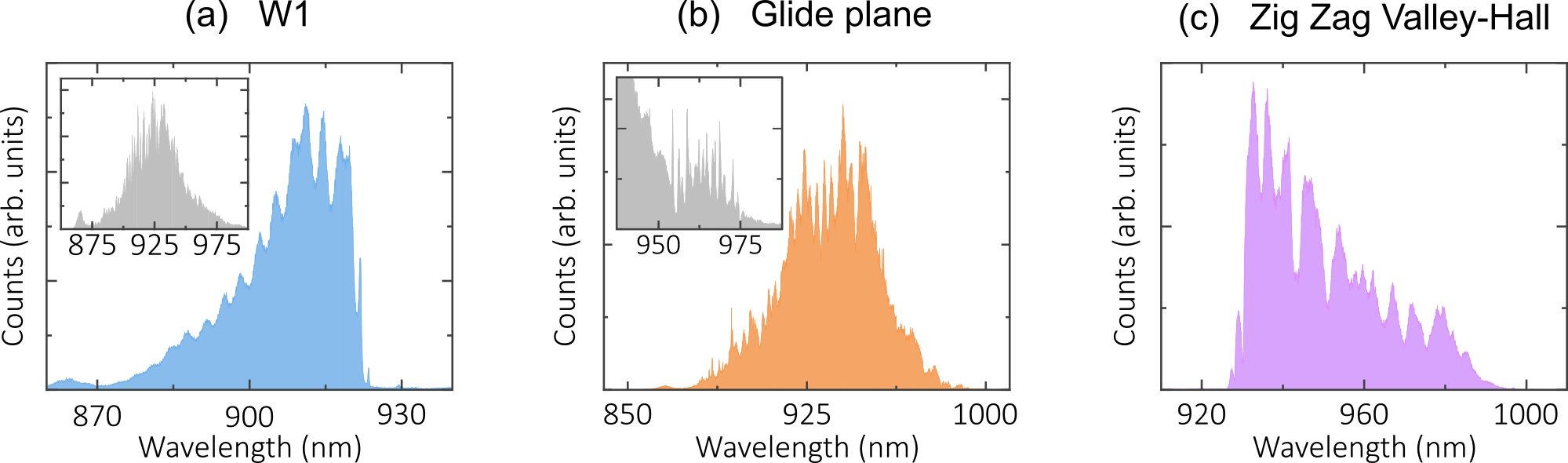}
\caption{\label{Fig2s} (a-c) Transmission of QD photoluminescence (PL) through an (a) W1, (b) glide-plane, and (c) VH waveguide, respectively. In each case, PL was generated using high-power non-resonant excitation of QDs located in a grating coupler at one end of the waveguide, and collected from the coupler at the opposite end of the device after transmission through the waveguide. The inset in (a) provides a reference PL spectrum for QDs in the bulk. The inset in (b) shows the PL signal measured from one outcoupler when exciting the QDs located in the glide-plane waveguide, revealing high finesse peaks within the slow light (and multimode) spectral window.}
\label{fp}
\end{figure*}

In this work, we identified the regions of the waveguides by considering both the transmission properties and the Fabry-P\'{e}rot (FP) fringes of the waveguide. By exciting the waveguide with either an above band light source to create broad QD emission, or the use of a broadband light source, we observed the FP fringes in the transmission spectrum. Fig \ref{fp} (a-c) shows transmission spectra for the W1, glide-plane and zig-zag VH waveguides. As expected, the spectra contain sharp peaks originated from the FP resonances sustained by the reflection at the waveguide ends. From the spacing between the fringes, we were able to determine the group velocity of the waveguide at different spectral points. 

Considering the case of the W1 waveguide, we can see a prominent drop in transmission at $\sim$ 920nm, an indication of the band edge. For the glide-plane waveguide, we observed high finesse modes in the wavelength range of $955-975$nm. This occurrence can be attributed to the overlapping slow light (flat band) spectral window for the two modes of the glide-plane waveguide. By identifying the regions of low and high group velocity, and locating the band edge from regions of low transmission we were able to work out the operational single mode bandwidths of the various waveguides from which we exclusively collected chiral data.

\newpage
\section{Calculation of $\beta$-factor}
\begin{figure}[h]
    \centering
    \includegraphics[width=0.99\textwidth]{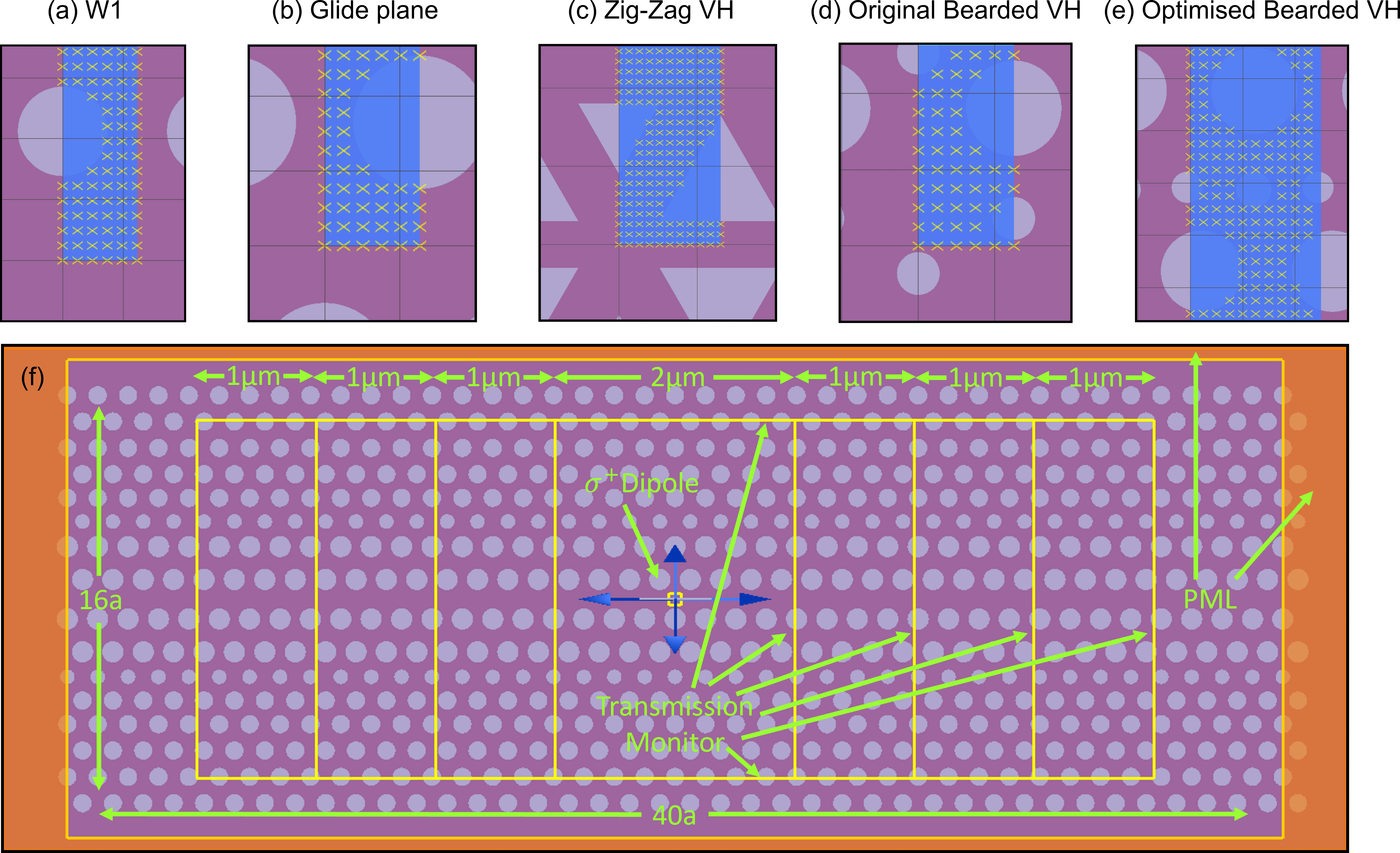}
    \caption{ (a)-(e) Detailed view of dipole positions (yellow crosses) used in calculating the positional and spectral dependence of the $\beta$-factor with regard to a) W1, b) glide-plane, c) zig-zag VH, d) original bearded VH and e) optimised bearded VH waveguide respectively. The irreducible repeating unit of each waveguide is shaded in blue. (f) Schematic of the FDTD $\beta$-factor simulations carried out. A dipole source with in-plane circular polarisation ($\sigma^{+}$) is surrounded by a series of transmission boxes that record the radiative power into the waveguide mode, in-plane loss, and out-of-plane loss.}
    \label{fig:my_label}
\end{figure}

The $\beta$-factor of a waveguide in the contect of chiral quantum optics, quantifies the coupling efficiency of a quantum emitter into cavity/waveguide mode. It is defined as the spontaneous emission rate into guided modes over total decay rate. That is,
\begin{equation}
\beta=\frac{\Gamma_{bound}}{\Gamma_{bound}+\gamma_{ng}}\:,
\end{equation}
where $\Gamma_{bound}, \gamma_{ng}$ are decay rates into a guided and non-guided mode respectively. The calculation of the $\beta$-factor for this work is done through finite-difference-time-domain (FDTD) simulations, using the the commercial FDTD software package, \textit{Lumerical} \cite{Lumerical}. The simulations consist of measuring the power throughput of a circularly polarised dipole emitter, collected from a series of transmission/power monitors surrounding the source. The $\beta$-factor is then calculated as the fraction of power coupled into the propagating modes over the total power injected from the emitter. Transmission monitors that account for out-of-plane and in-plane losses are placed two lattice constants above the waveguide and six lattice constants away from the waveguide centre respectively, while those for the waveguide modes are placed at 1\textmu m, 2\textmu m, 3\textmu m and 4\textmu m away from the dipole source for transmission comparison as well as uncertainty calculations, as is shown in Fig. \ref{fig:my_label}. The simulation region is $20a$ in $x$, $16a$ in $y$ and $5a$ in $z$, with $42, 8, 8$ layers of perfect matched layers (PML) at the x, y and z boundaries. The simulation is run with a mesh of 20 cells per lattice constant. A refractive index of $3.4$ and $1$ is used for the GaAs dielectric slab and air holes respectively. We vary the circularly polarised dipole's position within the least irreducible unit of the waveguide, up to $1.5a\sim 2a$ away from the waveguide centre (see Fig.\ref{fig:my_label} for more details with regard to each waveguide). 

During the $\beta$-factor calculations, we observed minor, high frequency fringes in the transmission data caused by weak reflection at the PML region and the finite sampling rate inherent to the nature of the Fourier transforms of FDTD. In order to remove the fringes, we apply a Savitzky–Golay Filter. The filter is applied to all three elements (i.e. forward, backward and total electric field power) used for the $\beta$ factor calculation.


\newpage
\section{Loss into free space modes at chiral points}

Since the $\beta$-factor characterises the fraction of spontaneous emission into waveguide propagating mode, it is therefore closely associated with the Purcell-factor, which describes the spontaneous emission rate enhancement of a dipole transition by its surrounding environment in the weak coupling regime. For waveguides the Purcell factor is proportional to the group index and the electric field intensity, and can be written as: \cite{beta}

\begin{equation}
F_P(r)=\frac{3\pi c^2a}{\omega^2\sqrt{\epsilon_r(r)}}\cdot\frac{1}{v_g}\cdot\frac{|\hat{e}_k(r)\cdot\hat{n}|^2}{\int\epsilon_r(r)|\hat{e}_k(r)|^2\rm{d^3}r}\:,
\end{equation}
where $\hat{e}_k(r)$ is Bloch mode profile at position $r$, $\hat{n}$ is a unit vector that denotes dipole orientation, $v_g$ is group velocity, $\epsilon_r$ is dielectric permittivity, $a$ is the period of the waveguide. The relation between Purcell Factor and $\beta$-factor is given by,
\begin{equation}
\beta=\frac{F_P}{F_P+F_r}\:,
\end{equation}
where $F_P$ is Purcell factor, $F_r$ is a factor representing the loss into free space modes, which takes a value between 0 and 1. By calculating both the $\beta$-factor as described in the previous section and the Purcell enhancement of the dipole sources at chiral points, we are able to extract the loss into free space modes at chrial points for each waveguide at the frequency point indicated by the red circles in Fig.1 in the main text. If the waveguide contained multiple locations where the chiral contrast was 1, we choose the point with the highest Purcell enhancement. For each waveguide we extract values for $F_r$ of:

\begin{table}[!ht]
    \centering
    \begin{tabular}{|l|l|}
    \hline
        Waveguide Type & $F_{r}$ \\ \hline
         W1 & $0.05\pm 0.01$ \\ \hline
        Standard GP & $0.05\pm 0.01$ \\ \hline
        Optimised GP &  $0.06\pm 0.01$ \\ \hline
        Zig-zag VH & $0.10\pm 0.02$ \\ \hline
        Original bearded VH &  $0.06\pm 0.02$ \\ \hline
        Optimised bearded VH &  $0.05\pm 0.01$ \\ \hline
    \end{tabular}
\end{table}


\newpage
\section{Spectral and spatial dependence of $\beta$-factor}
\begin{figure}[h]
    \centering
    \includegraphics[width=0.99\textwidth]{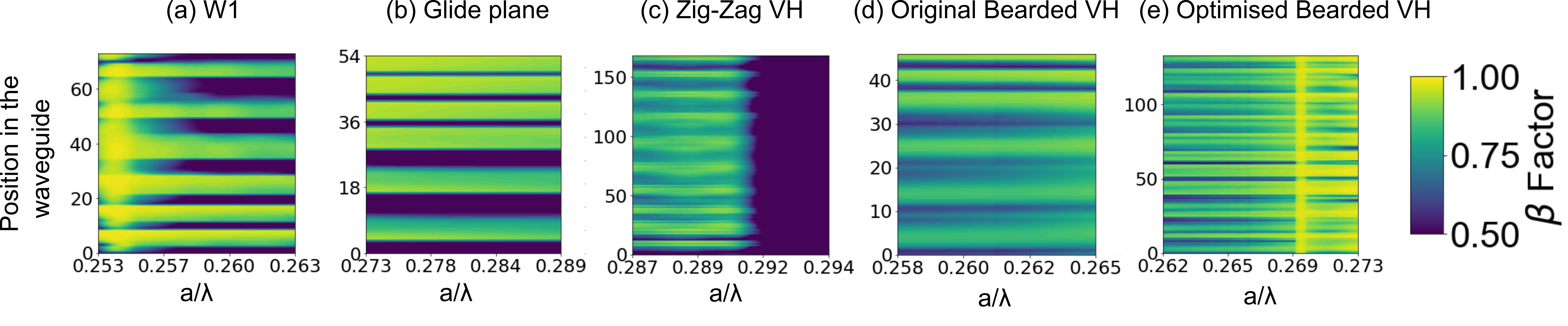}
    \caption{Frequency (horizontal-axis) and position (vertical-axis) dependence of $\beta$-factor for a) W1, b) glide-plane, c) zig-zag VH, d) original bearded VH and e) optimised bearded VH waveguide. See Fig.\ref{fig:my_label} (a-e) for detailed dipole positions. Range of the colour-map is set to $0.5\sim1$ so as to highlight the contrast.}
    \label{fig:my_label1}
\end{figure}

The spectral and position dependence of $\beta$-factor for the five waveguides explored in the main text is shown in Fig \ref{fig:my_label1}. Dipole positions are labelled first along the vertical axis (upward, away from the waveguide centre), then the horizontal axis (rightward, along the propagation direction). The frequency range for each waveguide is deduced from the single-mode regime of its band structure (see Fig. 1 in the main text). 

For the W1 waveguide, the single-mode region goes across the whole band, with a gradual decrease in group velocity, which reaches zero at the Brillouin Zone (BZ) edge. It has a near-invariant mode profile, contributing to a monotonic increase in the Purcell Factor against mode frequency. It is interesting to see a considerable $\beta$-factor in the presence of slow-light even at positions in the waveguide that have an absence of high electric field intensity. 

As for the standard glide-plane waveguide, its single-mode region only covers a small fraction of the upper band. Since there is no slow-light effect in this region and the electric field is well concentrated in the waveguide centre, the $\beta$-factor is lower than the W1 waveguide. Dispersion is approximately linear in the frequency range of interest, resulting in a near-constant $\beta$-factor across the single-mode region.

The Zig-Zag Topological VH waveguide is a single-mode waveguide. The modes possess a broad mode profile, with a mode waist radius of up to $3a\sim 6a$ depending on the relative contrast of two triangular holes \cite{taper}. Although this waveguide has a single mode slow light region, this region overlaps with the bulk crystal band and so can not confine the modes within the waveguide centre (see the dark region of short wavelength modes in Fig.\ref{fig:my_label1}). Furthermore, the broad mode profile makes the mode less concentrated, giving rise to low $\beta$-factors generally throught the waveguide's operation. 

Similar to the standard glide-plane waveguide, the original bearded VH waveguide has a slow-light region but unfortunately, once again it is multi-mode. Given that these modes tend to possess opposite polarisation, chirality is compromised and the slow-light region is no longer appropriate for chiral applications. With the additional consideration of its broad mode profile, the original bearded VH waveguide performs the worst in the $\beta$-factor comparison.

A significant improvement of the optimised bearded VH waveguide is the breaking of glide-plane symmetry which resolves the multi-mode issue by shifting the high-order mode upward. The introduction of a slow-light region away from the BZ edge brings about high Purcell enhancement for topologically-protected, gap guided modes with high electric field concentration. As a result, the optimised version of the topological Bearded VH outperforms all other waveguides explored in the main text in terms of $\beta$-factor .


\newpage
\section{QD chirality asymmetry}

\begin{figure}[h]
    \centering
    \includegraphics[width=0.75\textwidth]{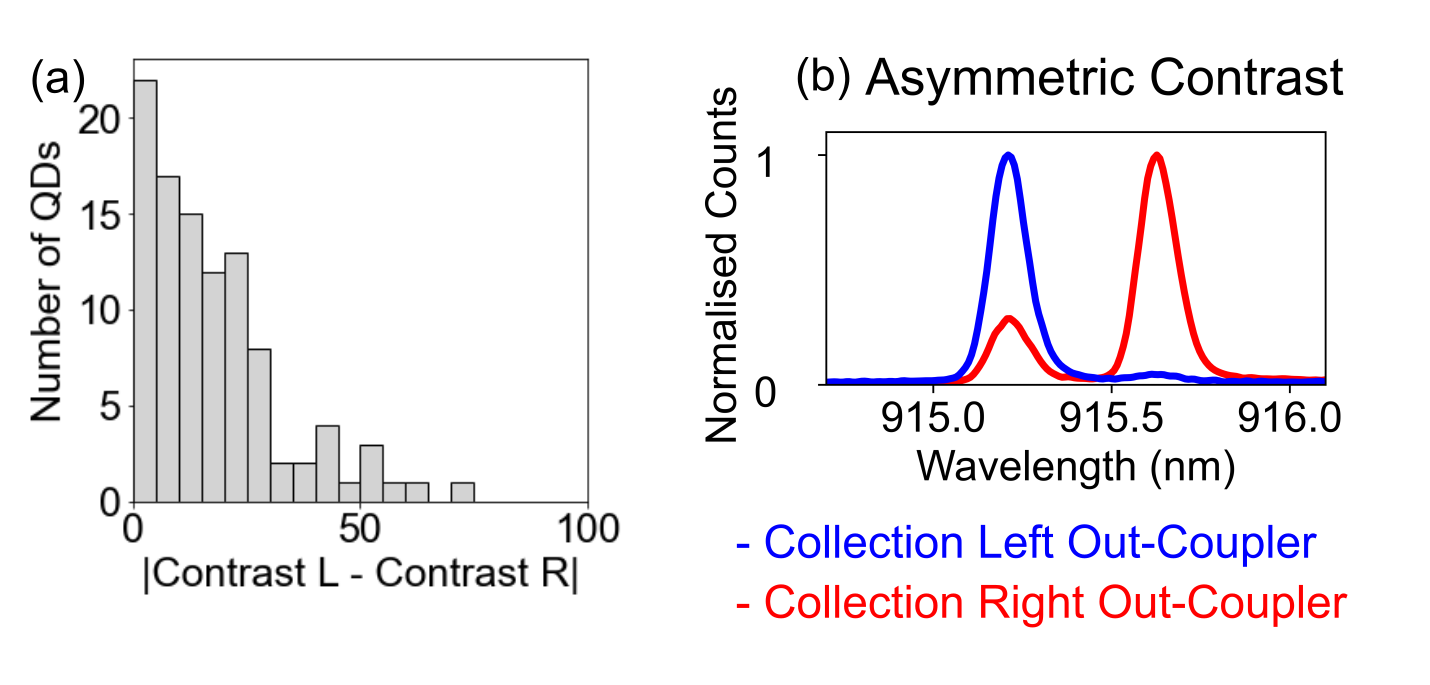}
    \caption{(a) Frequency of the asymmetry in the chiral contrast measured from the left and right out-coupler for the W1 waveguide. (b) spectra of an asymmetric QD, showing the difference in contrast viewed from the right out-coupler in red and the left out-coupler in blue.}
    \label{fig:S5}
\end{figure}

A notable characteristic observed across all waveguides, consistent with findings from other experiments and QD wafers\cite{Coles_thesis,Barik2020,JalaliMehrabad_Optica}, is the presence of asymmetry in the chirality of the QDs when comparing the measured chirality from the left and right out-couplers of the waveguide. In Fig 3 (f-i), we can observe examples of QD spectra that closely align with the predictions generated by the FDTD simulations illustrated in Fig 2 (j). These spectra exhibit the anticipated behavior, where the outputs from the right out-coupler (represented in red) and the left out-coupler (represented in blue) are symmetric, acting as mirrored versions of each other. They display the same chiral contrast but with opposite intensities. However, the spectra of Fig S5 (b) presents an instance of asymmetric behavior, where the chiral contrast observed from the left out-coupler surpasses that observed from the right out-coupler. Fig S5 (a) shows the distribution of the asymmetry, showing the number of QDs that were found within the W1 waveguide with a given difference between the contrast as calculated from the left out-coupler, and the contrast as calculated from the right out-coupler. This discrepancy in behavior warrants a thorough investigation into its possible origin and the implications it holds for the practical applications of these devices.


\newpage
\section{Slab Waveguide}

\begin{figure}[h]
    \centering
    \includegraphics[width=0.99\textwidth]{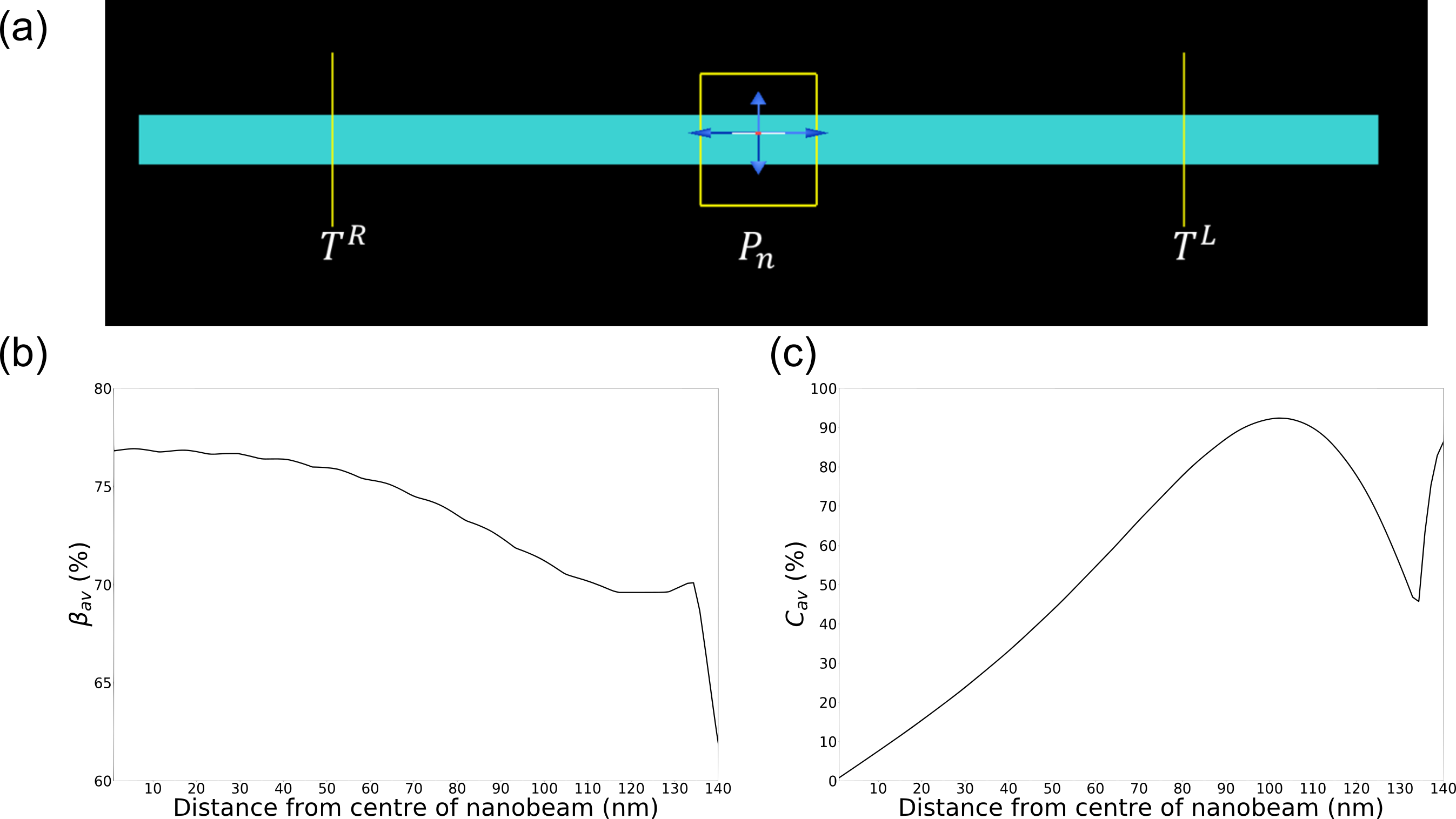}
    \caption{(a) Top down view of 3D-FDTD simulation with 2 transmission monitors $T^{R}$ and $T^{L}$  and a 3D optical power monitor $P_n$. (a) average chiral contrast and (b) average $\beta$-factor with respect to the dot position (starting in the middle of the waveguide at 0nm and moving towards the edge at 140nm). }
    \label{fig:S5}
\end{figure}

While the bulk of our work has focused on photonic crystal waveguides, for a complete comparison, we consider a simple slab waveguide. Here we analyse how the properties of chiral QD emission in a slab waveguide changes according to its position along the width of the structure. Using FDTD simulations and employing a circularly polarised dipole source, we can compare changes in average chiral contrast ($C_{av}$) and average $\beta$-factor ($\beta_{av}$) from normalised transmission monitors placed either side of the dipole ($T^{L}$ and $T^{R}$) with a 3D power monitor placed around the dipole source (layout shown in Fig S6 (a)).

With simulated results we can calculate $C_{av}$ and $\beta_{av}$  using:

\begin{equation}
     C_{av} = \frac{T^{R}_{av} - T^{L}_{av} }{T^{R}_{av} + T^{L}_{av}}  \;\;\; \text{and} \;\;\; \beta_{av} = \frac{T^{R}_{av} + T^{L}_{av}}{P_{norm}},
\end{equation}

where $T^{R}_{av}$ and $T^{L}_{av}$ is the mean transmission over the wavelength range 900-960 nm and $P_{norm}$ is the ratio of the injected power into the simulation region with the power that couples into the waveguide. To ensure the best possible coupling of a single fundamental mode into the waveguide, we restrict out width $w = 280$ nm (with a depth of 170 nm) in addition to mode expansion monitors which discard the minor coupling of higher-order modes.

From Fig S6 (c), we can see that circular dipole sources within the slab waveguide can show high contrast (max($C_{av}$) = 92\%), but the maximum $\beta$-factors possible are significantly lower than that of photonic crystals. 

\newpage
\section{Optimised Glide-plane}

\begin{figure}[h]
    \centering
    \includegraphics[width=0.99\textwidth]{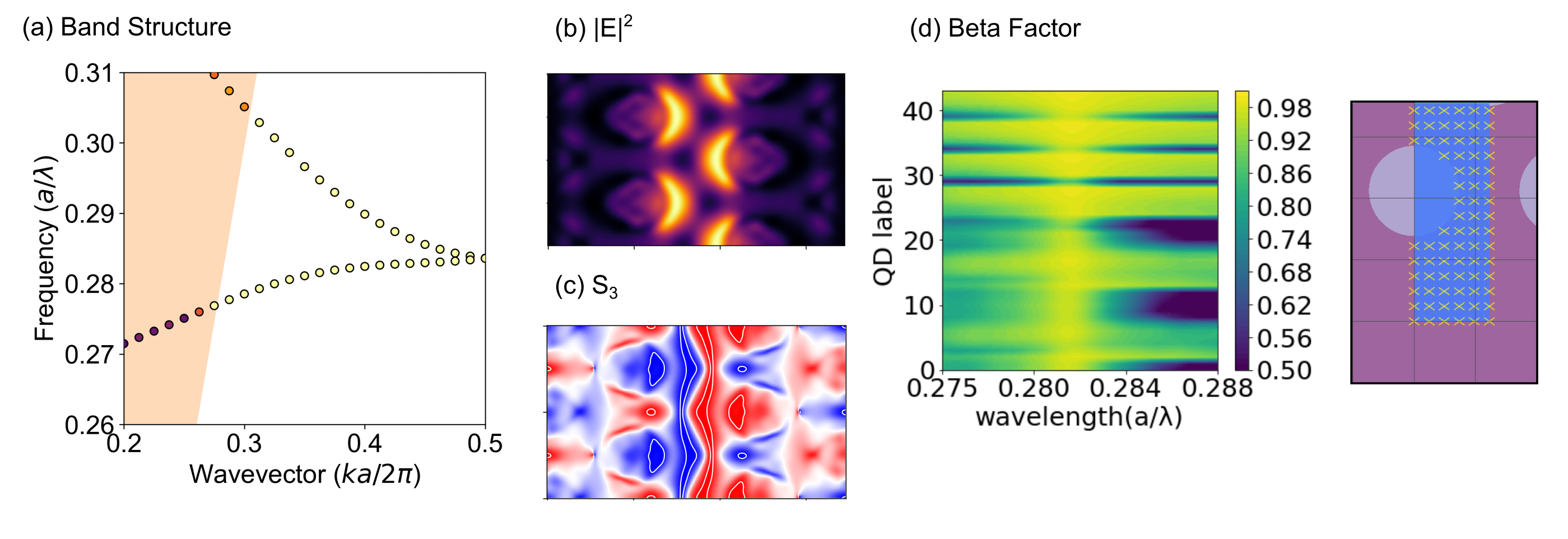}
    \caption{(a) Band structure for the optimised glide-plane wavetguides showing superated upper and lower guided modes, without a multi mode region. (b) The electric field intensity at the interface of the waveguide (c) The degree of circular polarisation of the electric field at the centre of the slab ($S_{3}$) (d) $\beta$-factor of the waveguide at different physics and spectral points in the waveguide.}
    \label{fig:S5}
\end{figure}

For the main text, the simplest versions of the conventional waveguides were used, for a fair comparison to the topological waveguide designs that haven't been optimised. However in this supplementary section we highlight a key candidate waveguide, the optimised glide-plane waveguide. Waveguides of this nature have featured predominately in work in quantum optics \cite{Hamidreza_glide}, \cite{lodahl_glide_plane}, with the design that is explored here coming from reference \cite{lodahl_glide_plane}. 

The optimisation of this waveguide, is mostly focused on creating a favourable band structure. The $S_{3}$ and electric field map of the waveguide interface is shown in Fig S7 (b-c), which is a very similar field profile as seen in Fig 2 (b) for the un-optimsied waveguide. 

In comparison to the un-optimised version, this waveguide has no multi mode region, and instead has single-mode operation from both guided modes of the waveguide. In the un-optimised waveguide, the slow-light region of operation, is also in a multi mode region. Because the two guided modes often have opposite circular polarisations of their electric fields for a single point at the interface of the waveguide, the multi mode region renders highly directional chiral emission impossible. 

The slow light region of the waveguide provides a significant improvement in the $\beta$-factor of the waveguide at all points within the interface. The drawbacks of the standard glide-plane waveguide design are addressed by the optimised version. The $\sim5nm$ wide slow-light region at BZ edge leads to high $\beta$ even if the emitter is placed $2a$ away from the centre. In figure S7 (d), we see a line of high $\beta$-factor at ~0.282 a/$\lambda$ corresponding to the slow light region.

\newpage

\end{document}